\documentclass[preprint,11pt,tightenlines,amsfonts,amssymb,amsmath,nofootinbib]{revtex4-1}

\usepackage{graphics}
\usepackage{graphicx}

\usepackage{setspace}
\usepackage{color}

\newcommand{\be}{\begin{equation}}
\newcommand{\ee}{\end{equation}}
\def\ud{\mathrm{d}}

\begin{document}

\title{Multiple choices of time in quantum cosmology}

\author{Przemys{\l}aw Ma{\l}kiewicz}
\email{pmalk@fuw.edu.pl}
\affiliation{National Centre for Nuclear Research, 00-681 Warszawa, Poland}
\affiliation{APC, Universit\'e Paris Diderot, Sorbonne Paris Cit\'e, 75205 Paris
Cedex 13, France}

\date{\today}

\begin{abstract}

It is often conjectured that a choice of time function merely sets up a frame for the quantum evolution of gravitational field, meaning that all choices should be in some sense compatible. In order to explore this conjecture (and the meaning of compatibility), we develop suitable tools for determining the relation between quantum theories based on different time functions. First, we discuss how a time function fixes a canonical structure on the constraint surface. The presentation includes both the kinematical and the reduced perspective, and the relation between them. Second, we formulate twin theorems about the existence of two inequivalent maps between any two deparameterizations, a {\it formal canonical} and a {\it coordinate} one. They are used to separate the effects induced by choice of clock and other factors. We show, in an example, how the spectra of quantum observables are transformed under the change of clock and prove, via a general argument, the existence of choice-of-time-induced semiclassical effects. Finally, we study an example, in which we find that the semiclassical discrepancies can in fact be arbitrarily large for dynamical observables. We conclude that the values of critical energy density or critical volume in the bouncing scenarios of quantum cosmology cannot in general be at the Planck scale and always need to be given with reference to a specific time function.\end{abstract}



\maketitle

\section{Introduction}
A time function is a key to the interpretation of quantum gravity. Although, its choice may be decided before or after quantization, the need for a time function at some point is inescapable \cite{Ku2}. But this choice is ambiguous, and as the evidence suggests, the quantum nature of the gravitational field is influenced by it. To our knowledge, the extent of this influence has never been studied thoroughly. Without resolving this issue, the quantum nature of the big bang cannot be properly addressed.
 
Perhaps the first ones to notice the multiple choice problem were Arnowit, Deser and Misner (see e.g. \cite{adm}), while the name which is most often associated with the problem is Karel Kuchar's. For a taste of his writing on this topic, see \cite{Ku}. The issue has appeared in a strictly quantum cosmological context in \cite{gotay0}, where the choice of time function was shown to determine the fate of classical singularity. The most basic fact about the time problem is that deparametrizations of hamiltonian constraint systems with respect to different time functions give canonically inequivalent reduced frameworks. Consequently, the imposition of quantum rules generically renders the models physically inequivalent.  It was found in \cite{Malkiewicz:2011sr} that different time functions lead to qualitatively distinct spectra for a Hubble rate operator in the Kasner universe: with respect to one time function the spectrum was continuous and unbounded, while with respect to another one it was discrete and bounded. Such discrepancies may be a reflection of unique physics entailed in the choice of time function. Some canonically inequivalent formulations have been also discussed in the context of a spherically symmetric dust shell \cite{kijowski}. A general observation pertaining to the problem of time in general relativity were made in \cite{hajicek}.

In the present work we are going to use the so called reduced phase space approach, which seems to us more suitable for this investigation. Nevertheless, the conclusion we reach are completely general. We are primarily concerned with comparing quantum theories with various variables playing the role of time. In order to do that we need to ensure that any discrepancies found should reflect a different choice of clock variable rather than a different choice of quantization and alike. Recall that in the light of the so called von Hove-Groenewald obstructions (see e.g. \cite{gotay}), there does not exist a faithful representation of many symplectic manifolds, including the most familiar one, the plane $\mathbb{R}^2$. As a consequence, quantizations are largely arbitrary, in particular they depend on the employed basic variables. 

The results obtained in this work are the following: We show that there exists a {\it formal canonical} mapping between any two deparameterizations, which (i) is different from a coordinate map and (ii) if turned into a unitary operator upon quantization, it ensures that the deparameterizations are based on a unique quantization. This makes the project for comparing different choices of clock feasible. Making use of the discrepancies between that map and a coordinate map, it is easy to show that discrepancies in any kind of semiclassical description must occur. The extent of the discrepancies are illustrated by an example, in which we show that they may be made arbitrarily large. This makes the maximal or minimal values of cosmological observables very sensitive to clocks. We also show that decompositions of Hilbert space given by self-adjoint operators corresponding to the same observable expressed with different clocks are in general very different. The application of the presented results to specific cosmological models is postponed until a next paper by present author. 

The paper is organized as follows. In Sec II we provide a basic introduction into the multiple choice problem supplemented with a discussion of some special cases of time transformations. Sec III is devoted to the central result, which is given by two (twin) theorems about the existence of the formal canonical relation between almost any two deparameterizations. We propose to make the relation a unitarity upon quantization to ensure that any quantum-level discrepancies are induced by a choice of time. The theorems are supplemented by an example. We also give a general argument to indicate the dependence of semiclassical trajectories on time function used to derive them. In Sec IV we prove that the discrepancies between semiclassical descriptions can be arbitrarily large. We conclude in Sec V.

\section{Multiple choice problem}

\subsection{ADM form of constraint}

Let $(X,P, \phi^r,\pi_s)$ be a symplectic chart of a neighborhood of the constraint surface $H\approx 0$ and let $X$ be a global time function. It suffices to focus on that part of the constrained surface, in which $\frac{dX}{d\tau}>0$, where $\tau$ is some external time parameter. Then, the constraint function $H$ satisfies $\frac{\partial H}{\partial P}=\frac{dX}{d\tau}>0$. Thus, via the implicit function theorem, $P$ at any point of the surface is a function of the rest of the canonical coordinates, i.e. $P=-h(X,\phi^r,\pi_s)$ for some $h$. In other words, $\tilde{H}:=P+h(X,\phi^r,\pi_s)\approx 0$ fixes the constraint surface unambiguously and $\tilde{H}$ is called the ADM form of the constraint $H$. Unfortunately, there seems to be  no general method for constructing the ADM forms of a given $H$ for a given choice of time function. Moreover, it is often the case that the coordinates $(\phi^r,\pi_s)$ occurring in the ADM forms have non-trivial ranges, which is sometimes recognized through the problem of reality of physical hamiltonian (or, of its self-adjointness at the quantum level). Furthermore, one usually has a weak control over the form of $h$ once the choice of time function has been made. However, as it will be shown, the choice of true hamiltonian and time function are in fact unrelated.

For discussions of the ADM form in the full general relativistic context, see Kuchar \cite{Ku1}, Isham \cite{Ish} and Hajicek and Kijowski \cite{Kij}. In the case of models with a single constraint in finite dimensional phase space, the discussion simplifies considerably. Suppose that there exists a canonical transformation from the initial canonical variables to new ones, in which the constraint $H$ takes the form:
\begin{equation}\label{H}
H=P+h(X,\phi^r,\pi_s)\approx 0
\end{equation}
where $(P,X)$ and $(\phi^r,\pi_r)$ are new canonical pairs. Then we observe that Hamilton's equations of motion for $X,\phi^r,\pi_s$ read:
\begin{equation}\label{red}
\dot{\phi}^r=\{\phi^r,H\}=\{\phi^r,h(X,\phi^r,\pi_s)\},~~
\dot{\pi}^s=\{\pi^s,H\}=\{\pi^s,h(X,\phi^r,\pi_s)\},~~
\dot{X}=1
\end{equation}
We construct an unconstrained formulation in the following way: we interpret the variable $X$ as an internal time variable and $h(X,\phi^r,\pi_s)$ as the generator of motion for $(\phi^r,\pi_r)$. The dynamical equations are now defined in the reduced phase space, which is parameterized by canonical pairs $(\phi^r,\pi_r)$. To distinguish it from the hamiltonian constraint $H$, $h(t,\phi^r,\pi_s)$ is called the true hamiltonian and has the meaning of the gravitational energy stored in $\phi^r$ and $\pi_s$ at $X=t$ hypersurface. The dynamics of the removed variable $P$ can be restored from (\ref{H}). Note that quantization of such a formulation follows the usual way leading to the corresponding Schr\"odinger equation.

The ADM form is of course not unique and there exist symplectomorphisms
\[(P,X,\pi_s,\phi^r)\mapsto (\tilde{P},\tilde{X},\tilde{\pi}_s,\tilde{\phi}^r)\] 
such that in terms of the new variables the constraint function can also admit the ADM form, i.e.:
\begin{equation}\label{2cf}
\tilde{H}=\tilde{P}+\tilde{h}_{\tilde{X}}(\tilde{X},\tilde{\pi}_s,\tilde{\phi}^r)\, ,
\end{equation}
where $H \approx \tilde{H}$. We included the index in $\tilde{h}_{\tilde{X}}$ to emphasize that the true hamiltonian generates the motion with respect to the internal variable $\tilde{X}$. Now, forming with Eq. (\ref{2cf}) the equations of motion like in Eq. (\ref{red}) and applying the following reduction of the full phase space to $(\tilde{\phi}^r,\tilde{\pi}_s)$, one arrives at another unconstrained formulation of the same physical system. As we will show shortly, the two unconstrained formulations will in general be \emph{canonically inequivalent}. 

Generally, the assignment of a time function, canonical variables and a true hamiltonian to a system, which is initially formulated as a hamiltonian constraint system, will be called deparameterization. 

For readers' convenience we recall the definition of canonical transformations as given in Abraham and Marsden \cite{AbMa}: Let $(\mathcal{P}_1, \omega_1)$ and
$(\mathcal{P}_2, \omega_2)$ be symplectic manifolds and
$(R\times\mathcal{P}_i, \tilde{\omega}_i)$ the corresponding
contact manifolds. A smooth mapping $F:R\times\mathcal{P}_1\mapsto R\times\mathcal{P}_2$ is called a {\it canonical transformation} if each of the following holds: (D1) $F$ is a diffeomorphism; (D2) $F$ preserves time; that is, $F^*t=t$; (D3) There is a function $K_F\in\mathcal{C}^{\infty}(R\times\mathcal{P}_1)$ such that $F^*\tilde{\omega}_2=\omega_K$, where $\omega_K=\tilde{\omega}_1+dK_F\wedge dt$. In what follows, the transformations which do not satisfy condition (D2) will be called non-canonical.

\subsection{Relation between kinematical and physical transformations}

\subsubsection{Time-preserving transformations}
For a class of kinematical symplectomorphisms restricted to the following form
\begin{equation}
X\mapsto \tilde{X}=X,~~P\mapsto \tilde{P}=P,~~\phi^s\mapsto \tilde{\phi}^s=\tilde{\phi}^s(\pi_r,\phi^t),~~\pi_s\mapsto \tilde{\pi}_s=\tilde{\pi}_s(\pi_r,\phi^t)
\end{equation}
all the respective reduced phase spaces will be related by the canonical transformation given simply by $(\phi^s,\pi_s)\mapsto(\tilde{\phi}^s,\tilde{\pi}_s)$ with the true hamiltonian, $h\mapsto\tilde{h}=h$, left unchanged. Actually, the more general symplectomorphisms\begin{equation}\label{t-pres}
X\mapsto \tilde{X}=X,~~~P\mapsto\tilde{P}=P+F(X,\pi_s,\phi^r),~~\phi^s\mapsto \tilde{\phi}^s=\tilde{\phi}^s(X,\pi_r,\phi^t),~~\pi_s\mapsto \tilde{\pi}_s=\tilde{\pi}_s(X,\pi_r,\phi^t)
\end{equation}
where $F$ is any real function, will be shown to correspond to time-dependent (for $dF\neq 0$) canonical transformations in the reduced phase space. In this case, the true hamiltonian is transformed as follows: $h\mapsto \tilde{h}=h+F$. We will specify the form of these transformations at the end of the present subsection. 

\subsubsection{General transformations}
The broader class of symplectomorphisms involve a transformation of variable X, 
\begin{equation}
X\mapsto\tilde{X}=\tilde{X}(P,X,\pi_s,\phi^r)~,
\end{equation}
on which the time parameter is based after the reduction. This class of kinematical symplectomorphisms renders, except for a very special case, canonically inequivalent transformations when viewed from the reduced phase space. Within this class, the following transformations\footnote{As pointed out by an anonymous referee, this is equivalent to say that the coordinate volume element $\ud V:=\ud P\ud X$ is preserved by the transformation, i.e. $\ud V=\ud\tilde{P}\ud\tilde{X}$.}
\begin{equation}
X\mapsto\tilde{X}=\tilde{X}(X,P),~~P\mapsto \tilde{P}=\int(\tilde{X}_{,X})^{-1}dP,~~\phi^s\mapsto \tilde{\phi}^s=\tilde{\phi}^s(\pi_r,\phi^t),~~\pi_s\mapsto \tilde{\pi}_s=\tilde{\pi}_s(\pi_r,\phi^t)
\end{equation}
are particularly interesting and they lead, as we will show below, to what we will define as \emph{almost canonically equivalent} reduced theories. Among them, a very special subclass is given by $X\mapsto\tilde{X}=\tilde{X}(X)$ with $\tilde{X}_{,X}\neq 0$. It is not canonical according to the definition given in Abraham and Marsden. There is a good reason, however, to include them as canonical in the study of hamiltonian constraint systems, because they do not change the canonical structure in the reduced phase space, which will become apparent later.

\subsubsection{Reduced phase space}
So far we have given the procedure of reducing the kinematical phase space via the ADM form of constraint and the associated Eqs (\ref{red}). Now we will explain it in different terms. It will make the above statements about canonical, almost canonical and non-canonical transformations in the reduced phase space more clear. In the kinematical phase space the Poisson bracket is well-defined and is equivalent to the symplectic form via $\{\cdot,\cdot\}=\omega^{-1}$. By the reduction of a hamiltonian constraint system we mean the following two-step procedure: First, one pulls back the kinematical symplectic form to the constraint surface $\omega|_{H=0}=E^*\omega$, where $E:\mathcal{C}\mapsto\mathcal{P}$ is the embedding of the constraint surface $\mathcal{C}$ in the kinematical phase space $\mathcal{P}$. The two-form $\omega|_{H=0}$ is closed and degenerate with the Hamiltonian vector field $v_H=\{\cdot,H\}$ being its null eigenvector. Therefore, the constraint surface does not admit any natural bracket structure, which normally is the inverse of a non-degenerate closed two-form. However, in order to obtain a reduced canonical framework one must construct such a structure. It is achieved by the choice of time function $X: \mathcal{C}\mapsto \mathbb{R}$ such that surfaces of constant $X$ intersect each physical orbit once and only once, i.e. $v_H(X)\neq 0$ everywhere. In the second step, one pulls back the degenerate two-form to a constant $X$ surface, in which it becomes non-degenerate and in which it may be inverted to a unique Poisson bracket. The bracket can be transported to the constraint surface and thus if we repeat this process for each value of $X$ we obtain a well-behaved Poisson bracket in the whole constraint surface \cite{Malkiewicz:2011sr}. This imposed Poisson bracket turns out to be uniquely given by the two properties: {\bf 1.} the imposed bracket between any two given Dirac observables is determined by the kinematical bracket between any two corresponding weak Dirac observables, and {\bf 2.} the imposed bracket between any Dirac observable and the time function $X$ vanishes.

\begin{figure}[t]
\begin{tabular}{cc}
\includegraphics[width=0.45\textwidth]{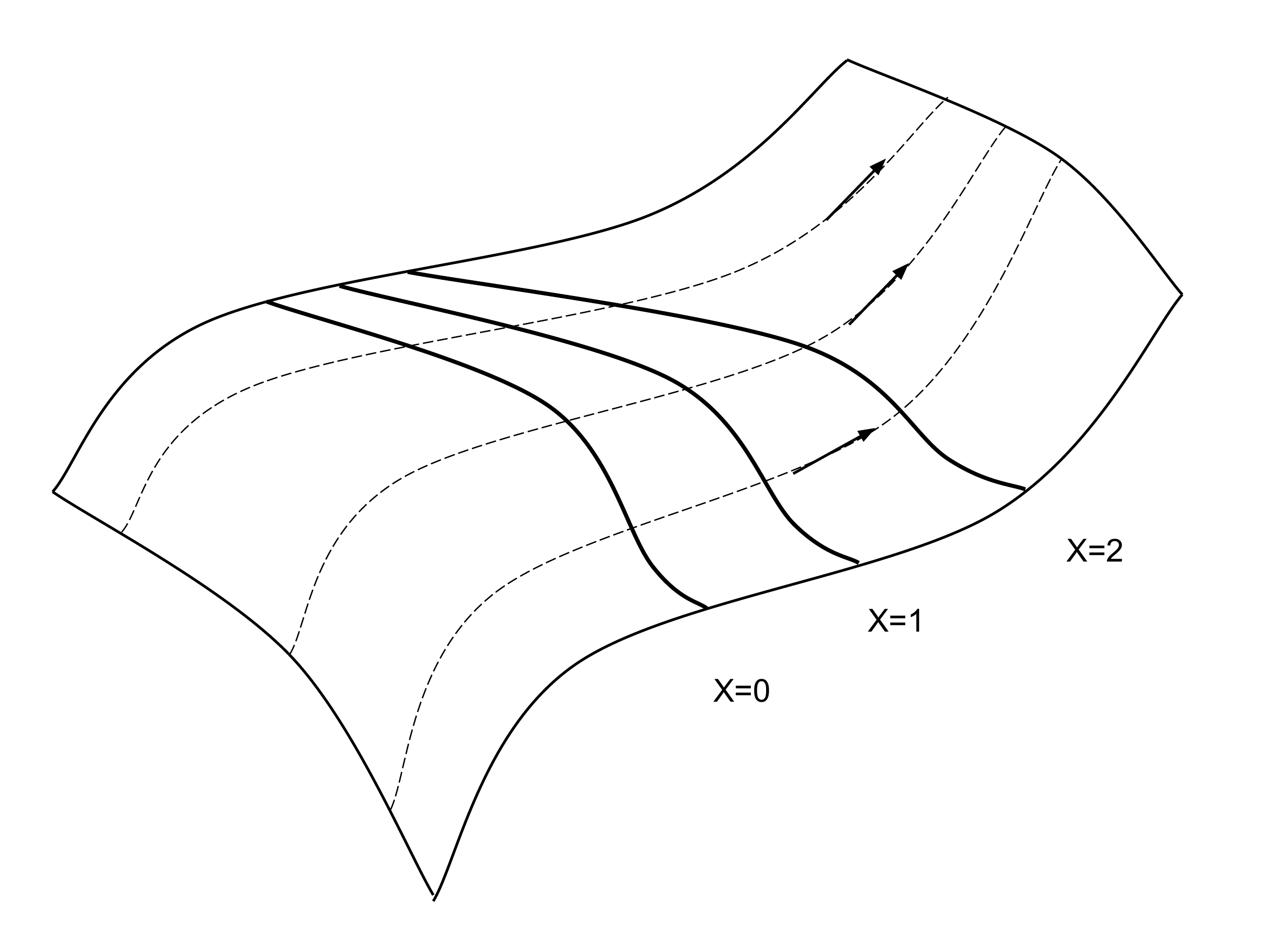}
\end{tabular}
\caption{\small The structure of the constraint surface: the orbits (dashed lines) generated by Hamiltonian vector field are crossed once and only once by each value of the clock variable (solid lines). The constant time submanifolds are embeddings of the physical phase space at the respective value of clock variable, where the Poisson structure is determined.}
\label{fig1}
\end{figure}

Expressing the above procedure in the kinematical variables $(P,X,\pi_s,\phi^r)$, which we used for the ADM form, we see that the first step is to obtain the reduced form:
\begin{equation}\label{redfor}
\omega|_{H=0}=\left(d\pi_rd\phi^r+dPdX\right)|_{P=-h_X(X,\phi^r,\pi_s)}=d\pi_rd\phi^r-dh_XdX
\end{equation}
In the second step, at each $X=const$ slice we invert $\omega|_{H=0,~X}$ to obtain:
\begin{equation}
\{\cdot,\cdot\}_X=\frac{\partial\cdot}{\partial\phi^r}\frac{\partial\cdot}{\partial\pi_r}-\frac{\partial\cdot}{\partial\pi_r}\frac{\partial\cdot}{\partial\phi^r}
\end{equation}
We see immediately that $\{X,\cdot\}_X=0$, which confirms the property no. {\bf 2.} of the imposed Poisson bracket, which was stated above. 

From the above construction it is easily seen that two brackets $\{\cdot,\cdot\}_X$ and $\{\cdot,\cdot\}_{\tilde{X}}$, which follow from two distinct constraint functions in Eq. (\ref{2cf}), will be the same if and only if $f(\tilde{X})=X$ for some (monotonic) $f$, i.e. if the two time functions give the same {\it slicing} of the constraint surface. It justifies calling such transformations \emph{canonical} in the reduced phase space. We easily obtain in this case that for the new time function $\tilde{X}$ (keeping the variables $\pi_r$ and $\phi^t$ unchanged) the degenerate two-form (\ref{redfor}) reads:
\begin{equation}\label{9}
\omega|_{H=0}=d\pi_rd\phi^r-dh_{\tilde{X}}d\tilde{X}
\end{equation}
where $h_{\tilde{X}}=f_{,\tilde{X}}h_{X}$, and
\begin{equation}\label{10}
\{\cdot,\cdot\}_{\tilde{X}}=\frac{\partial\cdot}{\partial\phi^r}\frac{\partial\cdot}{\partial\pi_r}-\frac{\partial\cdot}{\partial\pi_r}\frac{\partial\cdot}{\partial\phi^r},~~~\{X,\cdot\}_{\tilde{X}}=f_{,\tilde{X}}\{\tilde{X},\cdot\}_{\tilde{X}}=0
\end{equation}
Note also that if the true hamiltonian $h_X$ is time-independent, then the new true hamiltonian $h_{\tilde{X}}$ depends on time in a very simple way. If we quantize $h_{X}\mapsto\hat{h}_{X}$ and define a unitary evolution operator $U=e^{-i\hat{h}_{X}(X-X_0)}$, then the dynamics with respect to $\tilde{X}$ is automatically given by $\tilde{U}=e^{-i\int\hat{h}_{\tilde{X}}d\tilde{X}}=U$. Moreover, at the quantum level, the algebra of all quantum observables is exactly the same. Thus, time redefinitions, which preserve the slicing, do not really lead to any change at the quantum level either.

Now let us turn to more general transformations, which involve a change of time function of the following form: $X\mapsto\tilde{X}=\tilde{X}(X,h_X)$, where $\tilde{X}_{,h_X}\neq 0$. Let us call them \emph{almost canonical}\footnote{They cannot be called canonical in the sense of the definition given at the end of subsection II A.}. In other words, by an almost canonical transformation we mean such a transformation of time $X$ and canonical coordinates $(\pi_r,\phi^r)$ that the final time $\tilde{X}=\tilde{X}(X,h_X)$ is a function of the initial true hamiltonian and initial time, while the canonical coordinates are preserved. We emphasize that this definition depends on the initial true hamiltonian. Explicitly, upon such a transformation the two-form (\ref{redfor}) reads:
\begin{equation}
\omega|_{H=0}=d\pi_rd\phi^r-dh_{\tilde{X}}d\tilde{X}
\end{equation}
where we kept $\pi_r$ and $\phi^t$ unchanged, $h_{\tilde{X}}=\int (\tilde{X}_{,X})^{-1}dh_{X}$ and
\begin{equation}
\{\cdot,\cdot\}_{\tilde{X}}=\frac{\partial\cdot}{\partial\phi^r}\frac{\partial\cdot}{\partial\pi_r}-\frac{\partial\cdot}{\partial\pi_r}\frac{\partial\cdot}{\partial\phi^r},\end{equation}
We compute
\begin{equation}
\{X,\cdot\}_{\tilde{X}}=X_{,h_X}\{h_X,\cdot\}_{\tilde{X}}\neq 0.
\end{equation}
Therefore, $\{\cdot,\cdot\}_{\tilde{X}}\neq \{\cdot,\cdot\}_{{X}}$ and the transformation is not canonical in the reduced phase space. Nevertheless, the Poisson algebra of all observables $O(\pi_s,\phi^r)$, which are explicitly time-independent, in particular of the basic variables, is preserved by this transformation and the quantization of these observables may lead to the same quantum representation. A  difference will only occur in quantization of those observables, which explicitly depend on time. Note that a special subclass of these transformations is very simple to analyze, namely for $X=f(h_X)\tilde{X}+g(h_X)$, i.e. linear dependence between two clock variables. Then 
\begin{equation}
h_{\tilde{X}}=\int f(h_X)dh_X=F(h_X)\, ,
\end{equation} 
where $F$ is an antiderivative of $f$. In this case, the existence of a unitary evolution at quantum level in one clock variable implies the existence of a respective unitary evolution in the other clock. Moreover, the quantum dynamics will be essentially the same: the same eigenfunctions, strictly related spectra, only the pace at which the change takes place will vary from one eigenstate to another. A particular case is for $h_X=h_{\tilde{X}}$, $f=1$ and $g_{,h_X}\neq 0$. At first it may be thought that these two deparametrizations are canonically equivalent. However, this is not the case, because the Poisson brackets are not identical ($X_{,h_X}\neq 0$ for $\tilde{X}$ fixed) and upon any time-dependent transformation of basic variables $(\pi_s, \phi^r, \tilde{X})\mapsto (\tilde{\pi_s}, \tilde{\phi^r})$, the seeming equivalence is lost. Nevertheless, the above discussion justifies calling these transformations \emph{almost canonical}.

\begin{figure}[t]
\begin{tabular}{cc}
\includegraphics[width=0.95\textwidth]{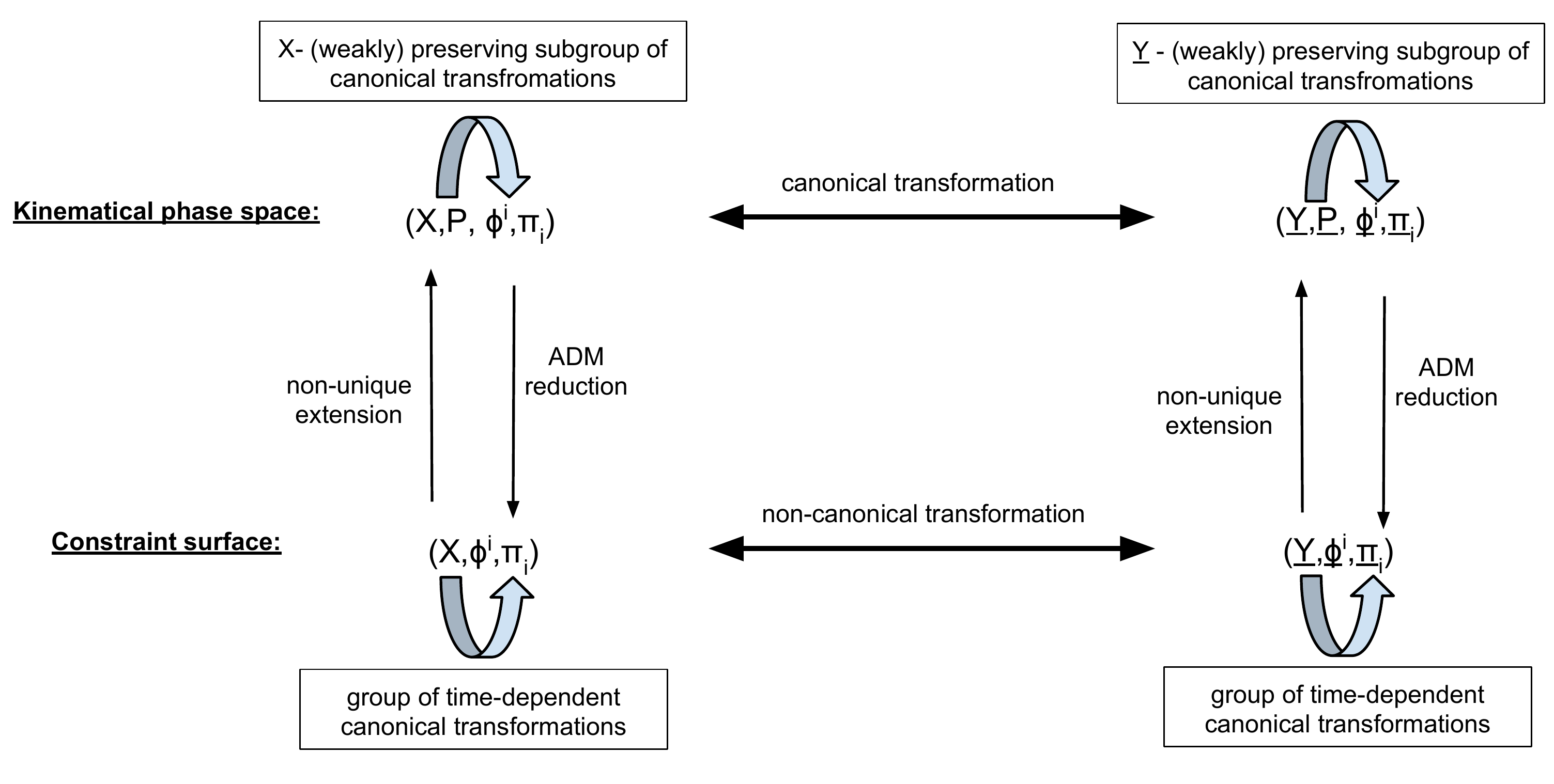}
\end{tabular}
\caption{\small The figure captures the relations between two different physical formulations of a constraint system, based on two time functions,  $X$ and $\underline{Y}$. To a general canonical transformation under which variable X is mapped into variable $\underline{Y}$, there corresponds a non-canonical transformation in the constraint surface. The weakly time-preserving canonical transformations in the kinematical phase space preserve time function in the constraint surface and correspond to canonical transformations there.}
\label{fig2}
\end{figure}

Let us consider now the special transformations, which do not change the time function. Suppose that $P\mapsto\tilde{P}=P+F(X,\pi_s,\phi^r)$ for some $F$ and let $\pi_s$ and $\phi^r$ remain unchanged for a moment, then the reduced form (\ref{redfor}) reads:
\begin{equation}
\omega|_{H=0}=\{d\pi_rd\phi^r-dF(X,\pi_s,\phi^r)dX\}-d\tilde{h}_XdX
\end{equation}
where $\tilde{h}_{X}=h_{X}-F$. The two-form in the bracket is not canonical, but it is closed and with a suitable choice of basic variables $(\pi_s,\phi^r, X)\mapsto (\tilde{\pi}_s,\tilde{\phi}^r)$ it can be brought to the canonical form such as:
\begin{equation}
\omega|_{H=0}=d\tilde{\pi}_rd\tilde{\phi}^r-d\tilde{h}_XdX
\end{equation}
A suitable transformation may be found with the help of Hamilton-Jacobi theory. We introduce a function $W(\phi^r, \tilde{\phi}^t, X)$, which generates the transformation via:
\begin{equation}\label{H-J}
\pi_rd\phi^r-F(X,\pi_s,\phi^r)dX - \tilde{\pi}_rd\tilde{\phi}^r=dW
\end{equation}
from which it follows that $W$ is required to satisfy the H-J equation:
\begin{equation}
F(X,\frac{\partial W}{\partial \phi^s},\phi^r)+W_{,X}(\phi^s, \tilde{\phi}^t, X)=0
\end{equation}
Now we can rewrite (\ref{t-pres}) as
\begin{eqnarray}
X\mapsto \tilde{X}=X,~~~P\mapsto\tilde{P}=P-W_{,X}(\phi^s, \tilde{\phi}^t, X),\\~~\phi^s\mapsto \tilde{\phi}^s=\left(W_{,\phi^s}(\phi^s, \tilde{\phi}^t, X)\bigg|_{\phi^s,X}\right)^{-1}\left(\pi_s\right),~~\pi_s\mapsto \tilde{\pi}_s=W_{,\tilde{\phi}^s}(\phi^s, \tilde{\phi}^t, X)
\end{eqnarray}
where the generating function $W$ must yield an invertible mapping between two sets of basic variables and therefore has to satisfy:
\begin{equation}
\textrm{det}(W_{,\phi^r\phi^s})\neq 0\neq \textrm{det}(W_{,\tilde{\phi}^r\tilde{\phi}^s})\end{equation}
We conclude that the Hamilton-Jacobi theory, which is essentially given by formula (\ref{H-J}), can be seen as a reduction of the theory of time-independent canonical transformations in a higher dimensional phase space, which preserve a fixed coordinate, and which were introduced with formula (\ref{t-pres}). The core findings of the present section are summarized in Fig. \ref{fig2}.

\section{Relation between deparameterizations}

\subsection{Complete time functions}
In what follows we discuss the basic connection between different dynamical deparameterizations of a hamiltonian constraint system, which will be useful for their comparison at the quantum level. We begin from the following two-form:
\begin{equation}\label{forsol}
\omega|_{H=0}=d\pi_rd\phi^r-dhdX,~~r\geqslant 1\, ,
\end{equation}
where we set $h=\pi_1$ and $X$ is a time function. The Poisson bracket reads:
\begin{equation}
\{\cdot,\cdot\}_{X}=\frac{\partial\cdot}{\partial\phi^r}\frac{\partial\cdot}{\partial\pi_r}-\frac{\partial\cdot}{\partial\pi_r}\frac{\partial\cdot}{\partial\phi^r}\end{equation}
Now, we consider a general transformation of clock function, $X=f(\tilde{X},\phi^r,\pi_r)$, without changing the true hamiltonian. We substitute $\tilde{X}$ into Eq. (\ref{forsol}) and obtain
\begin{equation}
\omega|_{H=0}=d\pi_1d({\phi}^1-f(\tilde{X},\phi^r,\pi_r)+\tilde{X})+d\pi_sd\phi^s-dhd\tilde{X},~~s\geqslant 2
\end{equation}
We introduce a new canonical variable ${\tilde{\phi}}^1={\phi}^1-f(\tilde{T},\phi^r,\pi_r)+\tilde{X}$ and obtain a new Poisson bracket:
\begin{equation}
\{\cdot,\cdot\}_{\tilde{X}}=\frac{\partial\cdot}{\partial{\tilde{\phi}}^1}\frac{\partial\cdot}{\partial\pi_1}-\frac{\partial\cdot}{\partial\pi_1}\frac{\partial\cdot}{\partial{\tilde{\phi}}^1}+\frac{\partial\cdot}{\partial\phi^s}\frac{\partial\cdot}{\partial\pi_s}-\frac{\partial\cdot}{\partial\pi_s}\frac{\partial\cdot}{\partial\phi^s},~~s\geqslant 2\end{equation}
Although these two reduced phase spaces based on separate time functions are canonically inequivalent as the basic variables ${\phi}^1$ and ${\tilde{\phi}}^1$ are physically distinct observables, they have the same formal structure including the same true hamiltonian, provided the ranges of ${\phi}^1$ and ${\tilde{\phi}}^1$ are the same. Therefore, \\
\\
\noindent {\bf Observation} {\it If one is successful at quantizing a gravitational system in one time variable then one is automatically successful at quantizing the system in all other time variables with the same range.
}\\
\\
The following considerations will generalize the above observation by showing how one relates any two deparametrizations so at the end only a single underlying quantum theory is needed. \\
\\
\noindent {\bf Lemma 1} {\it All the submanifolds in the constraint surface, which intersect each gauge orbit once and only once are mathematically equivalent as symplectic manifolds, and physically equivalent as the space of Dirac observables.
}\\
Proof. Let us show the equivalence between different submanifolds in the constraint surface: the reduced two-from, $\omega_{H=0}$, is invariant with respect to the action of any gauge transformation, which infinitesimally implies $\mathcal{L}_{\bar{t}}~\omega_{H=0}=0$, where $\bar{t}$ is any vector field tangential to the gauge orbits. Also, any two hyper-surfaces, which intersect all the orbits once and only once, are mapped into each other by some gauge transformation (generated by a suitable choice of $\bar{t}$). Thus, the symplectic form, which is the reduction of the two-form with respect to the hyper-surfaces must be {\it the same} under the gauge transformation in the both hyper-surfaces. 

\begin{figure}[t]
\begin{tabular}{cc}
\includegraphics[width=0.75\textwidth]{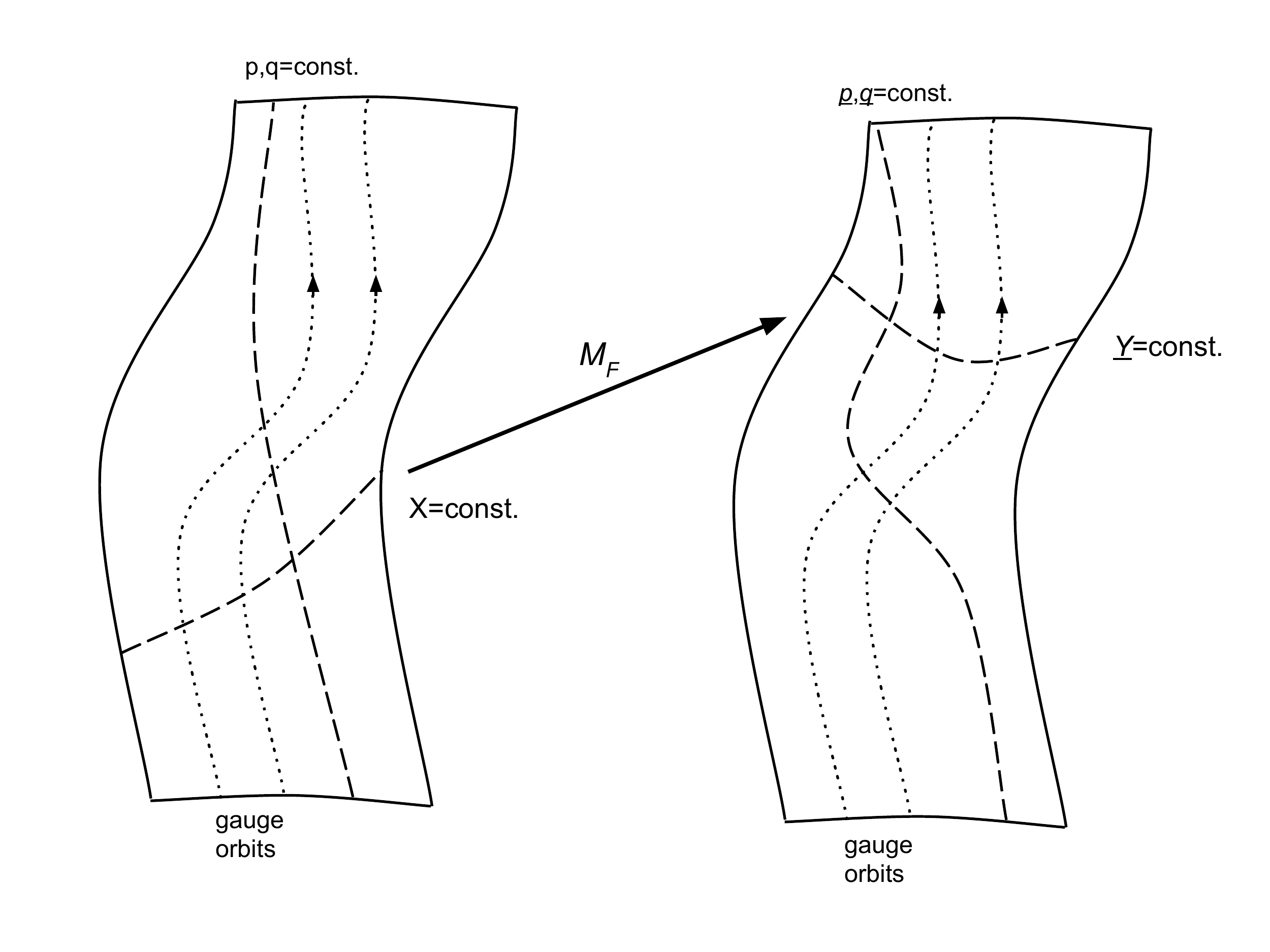}
\end{tabular}
\caption{\small The map $\mathcal{M}_F$ operates on the points in the constraint surface. It maps submanifolds of fixed value of time function $X$ into submanifolds of fixed value of time function $\underline{Y}$, where the submanifolds are represented by horizontal dashed lines. The gauge orbits, represented by the dotted lines, are invariant with respect to $\mathcal{M}_F$. The vertical dashed lines represent phase space variables $(q,p)$ and $(\underline{q},\underline{p})$, which are associated with a given deprametrization.} 
\label{fig3}
\end{figure}

Now, with the above lemma one may prove the following:\\
\\
\noindent {\bf Theorem 1} {\it Let $(X,\phi^r,\pi_s)$ and $(\tilde{X},\tilde{\phi}^r,\tilde{\pi}_s)$ be
two deparametrizations of a given hamiltonian constraint system and $h$ and $\tilde{h}$ be the respective true hamiltonians. Then for a given monotonic $F:\mathbb{R}\mapsto\mathbb{R}$, there exists a unique, invertible mapping  $\mathcal{M}_F:(X,\phi^r,\pi_s)\mapsto (\tilde{X},\tilde{\phi}^r,\tilde{\pi}_s)$ such that 
\begin{equation}X \mapsto \tilde{X}=F(X)~~~\textrm{and}~~~\pi\circ\mathcal{M}_F=\pi,\end{equation}
where $\pi$ is the projection generated by the gauge transformation. Moreover, the mapping satisfies:\\
\indent i) for every value of $X$, $\mathcal{M}_F^*|_{X}: (\tilde{\phi}^r,\tilde{\pi}_s)\mapsto (\phi^r,\pi_s)$ is canonical,\\
\indent ii) $dh=F'(X)d\mathcal{M}_F^{*}(\tilde{h})-i_{X}\mathcal{M}_F^*(d\tilde{\phi}^rd\tilde{\pi}_r)$}\footnote{By $i_{v}$ we denote the operation of contracting form with vector field $v$, e.g. $i_{v}\omega(\cdot,\cdot)=\omega(v,\cdot)$ for some two-form $\omega(\cdot,\cdot)$. Here, by $i_X$ we mean contraction with vector field $\partial_X$.}\\
Proof. Since $X=const$ and $\tilde{X}=const$ cross each gauge orbit once and only once, the extra condition $\pi\circ\mathcal{M}_F=\pi$ makes $\mathcal{M}_F$ uniquely defined. Then the point i) follows immediately from the lemma. Now we prove ii). Since $\pi\circ\mathcal{M}_F=\pi$, any vector that is tangent to a given orbit is mapped by $\mathcal{M}_F$ to a vector, which is tangent to the same orbit. In coordinates it reads:
\begin{equation}\mathcal{M}_{F*}:\partial_X+v_h\mapsto \lambda(X,\phi^r,\pi_s)\cdot (\partial_{\tilde{X}}+v_{\tilde{h}}),\end{equation}
where $v_h$ and $v_{\tilde{h}}$ are the hamiltonian vector fields and $\lambda(X,\phi^r,\pi_s)$ is an unspecified function. To specify $\lambda$ we note $\mathcal{M}_{F*}(\partial_X)=F'(X)\partial_{\tilde{X}}+\frac{\partial (\tilde{\phi}^r,\tilde{\pi}_s)}{\partial X}\partial_{(\tilde{\phi}^r,\tilde{\pi}_s)}$ and  $\mathcal{M}_{F*}(v_h)\cdot d\tilde{X}=0$, from which we obtain $\lambda=F'(X)$. This means that the Hamilton equations in $X$ are mapped into the Hamilton equations in $F^{-1}(\tilde{X})$. Since $\mathcal{M}_F^*(\omega_{H=0})=\omega_{H=0}$ we have $dXdh = \mathcal{M}_F^*(d\tilde{X}d\tilde{h})+\mathcal{M}_F^*(d\tilde{\phi}^rd\tilde{\pi}_r)-d\phi^rd\pi_r$. Upon contraction of the latter with $\partial_{X}$ we obtain ii). The setup of the theorem has been illustrated in Fig. \ref{fig3}.
\\

\noindent {\bf Remarks:} 

\noindent {\bf I.} Note that the theorem refers to {\it active} transformations in the constraint surface. This is in contrary to the Hamilton-Jacobi theory of time-dependent transformations, which are {\it passive}, that is, they are coordinate transformations on the constraint surface (and preserve time coordinate). More precisely, the transformation $\mathcal{M}_F$ is a combination of {\it passive} and {\it active} components, so that {\it a physical state} is mapped into {\it another physical state}, which is then parametrized with basic variables and time function of {\it different physical meanings}. To separate the active component of $\mathcal{M}_F$ one needs to know the relation between the tilde-accented and unmarked coordinates. Then the mapping splits into an active transformation, given as follows: 
\begin{equation}
(X,\phi^r,\pi_s)\mapsto (X',\phi^{r\prime},\pi_s')=\left(\pi\big|_{F(X)}\right)^{-1}\left(\pi(X,\phi^r,\pi_s)\right),\end{equation}
where $\pi\big|_{F(X)}$ is the invertible restriction of the natural projection to the submanifold of the constancy of $\tilde{X}=F(X)$, followed by the change of coordinate system represented by three functions: \begin{equation}\tilde{X}=\tilde{X}(X',\phi^{r\prime}\pi_s'),~~\tilde{\phi}^r=\tilde{\phi}^r(X',\phi^{r\prime},\pi_s'),~~\tilde{\pi}_s=\tilde{\pi}_s(X',\phi^{r\prime},\pi_s').\end{equation} We will denote this coordinate transformation by $\mathcal{R}:(X,\phi^r,\pi_s)\mapsto (\tilde{X},\tilde{\phi}^r,\tilde{\pi}_s)$.

\noindent {\bf II.} By the use of the theorem we may easily analyze the example considered above. There, the two time functions have the same range so that we could simply set 
\begin{equation}\label{M} (X,\phi^r,\pi_r)\mapsto(\tilde{X},\tilde{\phi}^r,\tilde{\pi}_r):=(X,\phi^r,\pi_r),~~r\geqslant 1\end{equation}
where we stress that by the equality `$:=$' we mean the numerical equality between values of coordinates in the new and old coordinate systems (the active component). The coordinate systems are related as follows (the passive component):
\begin{equation}\label{C1}\tilde{\pi}_1=\pi_1,~~\tilde{\phi}^s=\phi^s,~~\tilde{\pi}_s=\pi_s,~~s\geqslant 2\end{equation}
while  
\begin{equation}\label{C2}\tilde{\phi}^1=\phi^1-X+\tilde{X}. \end{equation}
We observe from (\ref{M}) that the transformation $\mathcal{M}_F|_{X}: (\phi^r,\pi_r)\mapsto(\tilde{\phi}^r,\tilde{\pi}_r)$ is canonical with
\begin{equation}
\{\phi^r,\pi_s\}_{X}=\delta^r_{~s}=\{\tilde{\phi}^r,\tilde{\pi}_s\}_{\tilde{X}}, 
\end{equation}
where all the remaining commutators vanish. On the other hand, the transformation given by (\ref{C1},\ref{C2}) is not canonical since e.g.
\begin{equation}
\{\phi^1,{\pi}_1\}_X=\{\tilde{\phi}^1+X-\tilde{X},\tilde{\pi}_1\}_X=\{\tilde{\phi}^1,\tilde{\pi}_1\}_X-\{\tilde{X},\tilde{\pi}_1\}_X\neq \{\tilde{\phi}^1,\tilde{\pi}_1\}_X
\end{equation}
This example nicely illustrates the general property that all clock deparametrizations are indeed related.

\noindent {\bf III.} We observe that $X$ and $\tilde{X}$ may have different ranges, and thus $F$ will not be the identity in general. But we have already noticed that time transformations of the form $X\mapsto F(X)$ preserve the canonical structure in the constraint surface and thus the choice of $F$ is irrelevant in this respect. As noticed in the discussion around Eqs (\ref{9}) and (\ref{10}), any finite clock formulation is equivalent to some infinite clock formulation. One of the consequences is that the range of clock alone cannot be of any relevance in quantum cosmology. In the next subsection we will generalize the concept of time function in such a way that a special class of time functions useful for deparametrizing singular systems will emerge.

\noindent {\bf IV.} From the theorem it immediately follows that the choice of time function and of the true hamiltonian are independent choices. Once the time function is fixed, the choice of basic variables fixes the true hamiltonian.

\noindent {\bf V.} We may use the theorem to study multiple choice problem. Namely, when we quantize two different deparameterizations of a hamiltonian constraint system and compare their physical implications we need to make sure that any discrepancies found are due to different choices of time rather than different quantum representations of the algebra of observables. This can be ensured in the following way:  At each $X$ (or $\tilde{X}=F(X)$) the variables $\phi^r$ and $\pi_s$ (or, $\tilde{\phi}^r$ and $\tilde{\pi}_s$) are in fact parametrizing the space of Dirac observables and at the same time they are joined by a certain canonical mapping $\mathcal{M}|_{X}: (\phi^r,\pi_s)\mapsto (\tilde{\phi}^r,\tilde{\pi}_s)$. If this mapping is promoted into a unitary operator upon the quantization, then both quantized deparametrizations are based on the same quantum representation of the Dirac observables. Notice that if both $h$ and $\tilde{h}$ are assumed to generate unitary transformations, the choice of value of $X$ (and $F(X)$) for $\mathcal{M}|_{X}$ is irrelevant. We will return to this point in Subsections \ref{C} and \ref{D}.

We need to remember that the above observation provides us with a {\it formal} correspondence stemming from the fact that all the imposed brackets must coincide in the space of Dirac observables and the Dirac observables parametrize the physical phase space at each instant of time completely. Nevertheless, the brackets differ in the space of dynamical observables and therefore the correspondence remains only formal, that is, a given variable will in general (i.e. if it is not a Dirac observable) have different physical meaning in each reduced phase space. Hence, if we have at hand a quantum theory derived in one time function then we may in principle use it to study other time-based quantum theories just by changing the physical interpretation of dynamical variables and their respective operators. This, of course, may have an effect in particular on the perceived singularity resolution scenario.

\subsection{Incomplete time functions}
So far we have assumed that the surfaces of constant time function cross each orbit once and only once. This assumption generically holds for dynamically complete systems with time parameter ranging from $-\infty$ to $+\infty$, but in gravitational systems this assumption is often violated, because: (i) the systems have no absolute clock, and there is no restriction on the range of internal clocks, (ii) the singular systems are dynamically incomplete with different orbits reaching the singular point at different physical conditions, and some of the internal clocks will take various values there. Note that the essential condition of monotonicity of time function along each orbit is preserved independently of (i) and (ii).

When we take the above point of view then each constant time surface crosses in general only a subset of orbits. This may be the case if some of the orbits have already terminated for a given value of clock. If so, then Lemma 1 does not apply. The natural question is if a relation analogous to the one described by theorem 1 exists among those more general deparameterizations. Let us formulate\\
\\
\noindent {\bf Lemma 2} {\it All the submanifolds in the constraint surface, which intersect an open subset of gauge orbits once and only once are mathematically equivalent as symplectic manifolds, and physically equivalent as subspaces of Dirac observables.
}\\
Proof: The proof is essentially the same as previously with the restriction of the constraint surface to an open subset of orbits.

Now, we wish to compare two deparameterizations with time functions such that the respective constant time submanifolds cross once and only once perhaps only a limited set of orbits. For convenience, we assume that only one of time functions is of the mentioned form while the other one crosses each orbit once and only once. We have the following:\\
\\
\noindent {\bf Theorem 2} {\it Let $(X,\phi^r,\pi_s)$ and $(\tilde{X},\tilde{\phi}^r,\tilde{\pi}_s)$ be
two deparametrizations of a given hamiltonian constraint system and $h$ and $\tilde{h}$ be the respective true hamiltonians. Suppose that constant $\tilde{X}$ surfaces cross each orbit once and only once while constant $X$ surfaces cross any orbit at most once. Then for a given monotonic $F:\mathbb{R}\mapsto\mathbb{R}$, there exists a unique injective mapping  $\mathcal{M}_F:(X,\phi^r,\pi_s)\mapsto (\tilde{X},\tilde{\phi}^r,\tilde{\pi}_s)$ such that 
\[X \mapsto \tilde{X}=F(X)~~~\textrm{and}~~~\pi\circ\mathcal{M}_F=\pi,\]
where $\pi$ is the projection generated by the gauge transformation. Moreover, the mapping satisfies:\\
\indent i) for every value of $X$, $\mathcal{M}_F^*|_{X}: (\tilde{\phi}^r,\tilde{\pi}_s)\mapsto (\phi^r,\pi_s)$ is canonical,\\
\indent ii) $dh=F'(X)d\mathcal{M}_F^{*}(\tilde{h})-i_{X}\mathcal{M}_F^*(d\tilde{\phi}^rd\tilde{\pi}_r)$}\\
\\
Proof. The proof, based on Lemma 2, is essentially the same as for Theorem 1. \\

\begin{figure}[t]
\begin{tabular}{cc}
\includegraphics[width=0.75\textwidth]{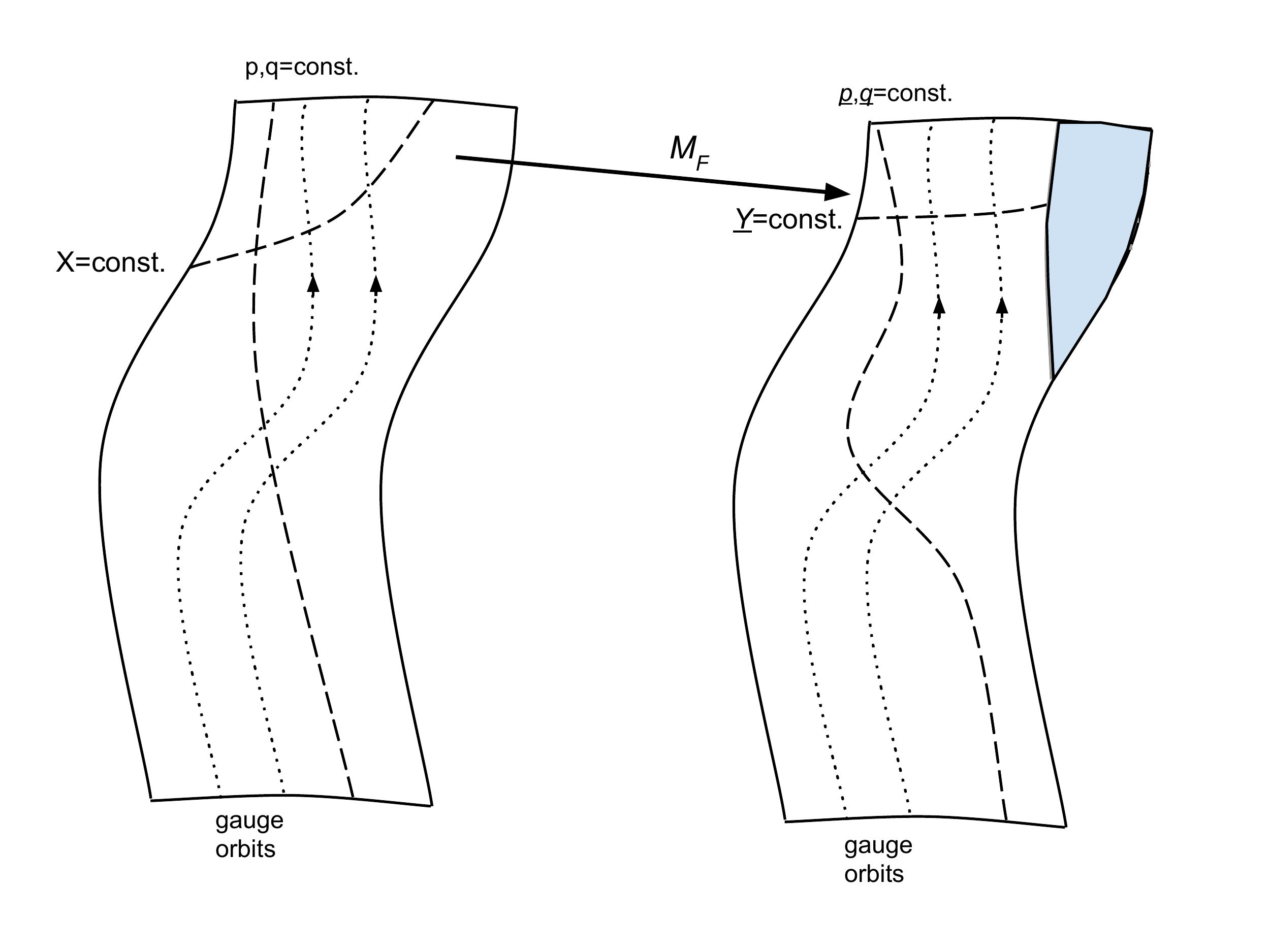}
\end{tabular}
\caption{\small The description of Fig \ref{fig3}.  also applies here. Since the time function on the left is incomplete and the one on the right is complete, there appears a dark region on the right, which is not covered by the image of the map $\mathcal{M}_F$. The latter is just another facet of the fact that the map $\mathcal{M}_F$ is not physical but formal.}
\label{fig4}
\end{figure}

\noindent {\bf Remarks:} 

\noindent {\bf I.} Note that even though each constant $X$ submanifold corresponds to a subspace of Dirac observables, in order for the parametrization to be complete, for each orbit there must exist such a value of $X$ for which the corresponding constant-time submanifold crosses this orbit once. Therefore we have in a sense complete correspondence between the Dirac observables in both parametrizations.

\noindent {\bf II.} The mapping $\mathcal{M}_F$ is injective but, in general, not surjective. This does not mean that the parametrization given on constant $X$ is incomplete in the sense that it does not cover all of the physical states. It does by assumption and the non-surjectivity is due to the active component of the mapping $\mathcal{M}_F$: it maps one-to-one the whole space of physical states into its subspace.

\noindent {\bf III.} Note that  in the parametrization $(X,\phi^r,\pi_s)$ for each orbit there exists a value of time function $X_0$ for which the orbit leaves the phase space, that is, the orbit is no longer crossed by the constant $X$ submanifolds for $X\geqslant X_0$. Therefore, the hamiltonian $h$ generates canonical transformation locally but not globally. In other words, the dynamics on each orbit is incomplete. This makes the time function $X$ attractive for modeling the gravitational singularity. Note that because $h$ does not generate a global canonical transformation, in general there will be no corresponding unitary transformation at the quantum level and the problem of singularity cannot be avoided simply by quantization. An extra input is necessary.

\noindent {\bf IV.} We can use Theorem 2 to study the multiple choice problem in the following way. As already pointed out, to make sure that any discrepancies between quantum theories with different time functions are due to the difference in time functions rather than quantum representation, we want to ensure that both theories are based on the same representation of Dirac observables. For each $X$, $\mathcal{M}|_X$ is a canonical map on a subset of the Dirac space and for this subset it should be ensures that the corresponding quantum operator is unitary. Unlike in Theorem 1, we have to ensure that such a corresponding unitary operator exist for each $X$ separately as there is no, in general, unitary transformation corresponding to $h$ to ensure unitarity between all pairs of constant-time slices once such a relation has been established for one of them.

\subsection{Quantum relations}\label{C}
In passing to quantum theory, each deparamterization will be assigned a space of linear operators on some Hilbert space. Let us denote two Hilbert spaces as $\mathcal{H}_X$ and $\mathcal{H}_{\tilde{X}}$, where $X$ and $\tilde{X}$ are some time functions. Then, just as proven, there exists a unique canonical map from deparametrization $X$ to deparametrization $\tilde{X}$ such that it preserves the space of Dirac observables and $\tilde{X}=F(X)$. We denote it by $\mathcal{M}_F$. Suppose it is turned into a unitary operator $\mathbb{M}:~\mathcal{H}_X\mapsto \mathcal{H}_{\tilde{X}}$. Then the relation between evolution operators reads:
\begin{equation}
U(X+\Delta X,X)=\mathbb{M}(X+\Delta X)^{\dagger}U(\tilde{X}+\Delta \tilde{X},\tilde{X})\mathbb{M}(X)
\end{equation}
and the subsequent relation between their generators, the true hamiltonians, follows
\begin{equation}\label{tru_ham}
\hat{h}_X=\mathbb{M}^{\dagger}(X)\left(\frac{d\tilde{X}}{dX}~\hat{h}_{\tilde{X}}+i\frac{d\mathbb{M}(X)}{dX}\mathbb{M}^{\dagger}(X)\right)\mathbb{M}(X)
\end{equation}
We emphasize that the existence of unitarity $\mathbb{M}$ is crucial for the two deperameterizations to be well-suited for comparison at quantum level. We note that if $\mathbb{M}$ is time-dependent, then true hamiltonians are not related by the unitary map $\mathbb{M}$, but an extra term appears. The relation (\ref{tru_ham}) supplemented with physical relation between the two clocks becomes a well-defined mapping between the respective Schr\"odinger equations. It is obvious that all the remaining operators are related as follows
\begin{equation}\label{trans}
\hat{O}_X= \mathbb{M}^{\dagger}\hat{O}_{\tilde{X}}\mathbb{M}
\end{equation}
Thus, any two theories are formally the same, perhaps except for their true hamiltonians. However, each operator has different physical interpretation depending on the choice of time. 

Now, we may be interested in if and how the choice of time affects the spectral properties of the operator corresponding to a fixed classical observable. The operators corresponding to observable $O$ in two different deparameterizations read then:
\begin{equation}
O\mapsto\hat{O}(\tilde{X})~\textrm{and}~O\mapsto\hat{O}(X)=\widehat{\mathcal{R}^{*}O(\tilde{X})}\end{equation}
where $\mathcal{R}$ is the coordinate map such that the pullback $\mathcal{R}^*({O}(\tilde{X}))={O}(X)$. The operators to compare are, say, $\hat{O}(X)$ and $\mathbb{M}^{\dagger}\hat{O}(\tilde{X})\mathbb{M}$, but since $\mathbb{M}$ is unitary for all times, it can be dropped. So, in fact, in this way we investigate the amount of non-unitarity indunced by the coordinate map, $\mathcal{R}$. The result clearly depends on the particular quantization map, which we employ and which needs to include many, perhaps non-trivial, compound observables on the phase space. 

\begin{figure}[t!]
\begin{tabular}{cc}
\includegraphics[width=0.7\textwidth]{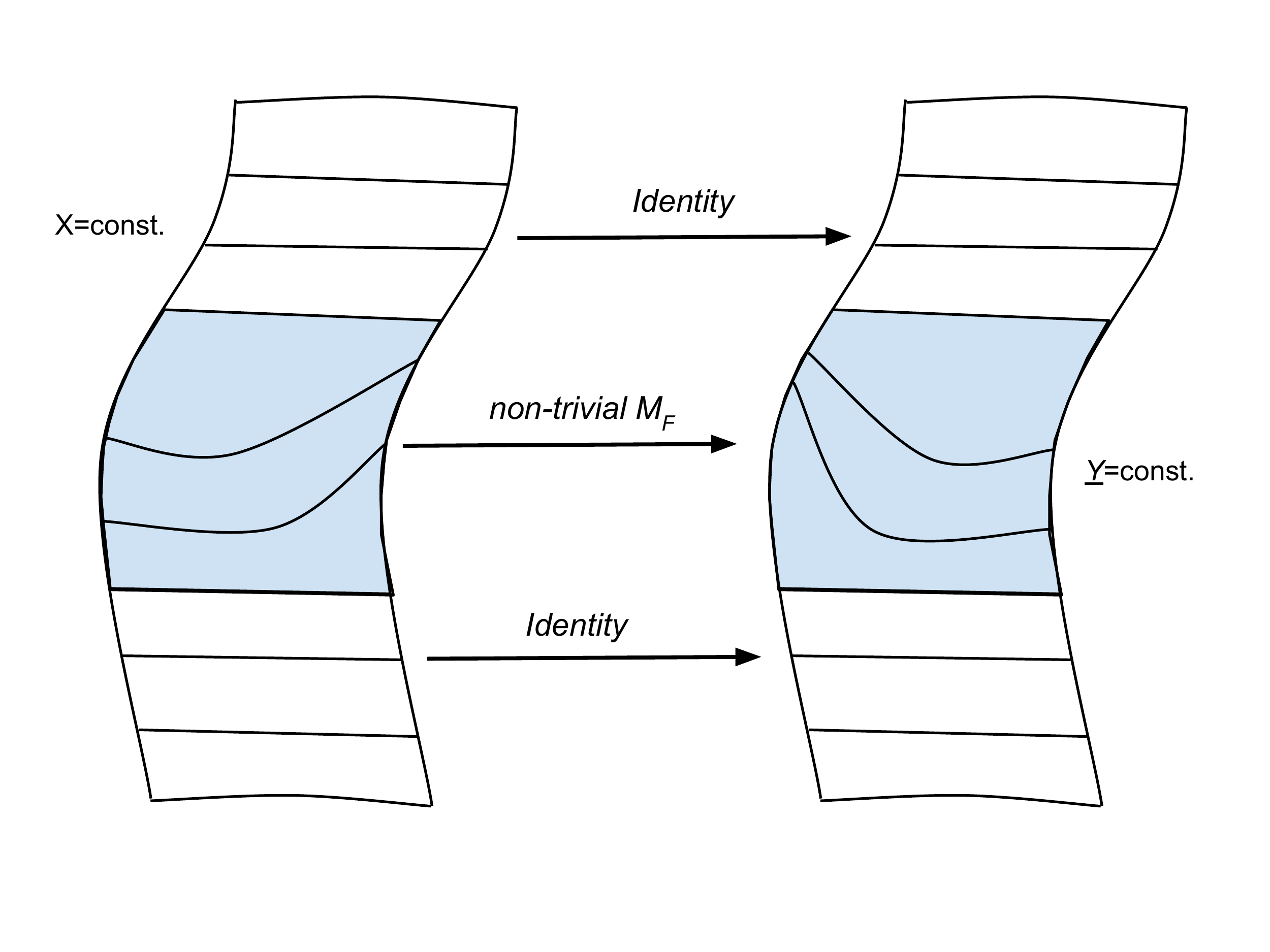}
\end{tabular}
\caption{\small  A sketch of map $\mathcal{M}_F$, which is not the identity only in a limited region of the constraint surface.}
\label{fig5}
\end{figure}

Alternatively, one may fix an operator and ask about its allowed interpretations, which come with possible choices of time function. That is, a fixed operator $\hat{O}(X)$ may be associated with the classical observable ${O}$ as well as with $\tilde{O}=\mathcal{R}^*\circ\mathcal{M}_{F(X)}^{-1*}({O})$. The advantage here is that the analysis is completely classical. The disadvantage is that one is unable to determine how quantum properties like spectra are affected by a change of time.

\subsection{Consequences}

Now, suppose that the constraint surface and the map $\mathcal{M}_F$ can be divided into three regions corresponding to, say, small, intermediate and large values of $X$. In region I,  $\mathcal{M}_F$ is the identity (its active part), in region II it differs from the identity significantly, and in region III it is the identity again. Moreover, we assume that quantum effects play a significant role in the evolution of the system in region II. Our considerations may refer to the big bounce scenario of quantum cosmology, with region I being the contracting branch, region II - the big bounce phase and region III - the expanding branch. The model is depicted in Fig. \ref{fig5}.

\begin{figure}[t]
\begin{tabular}{cc}
\includegraphics[width=0.7\textwidth]{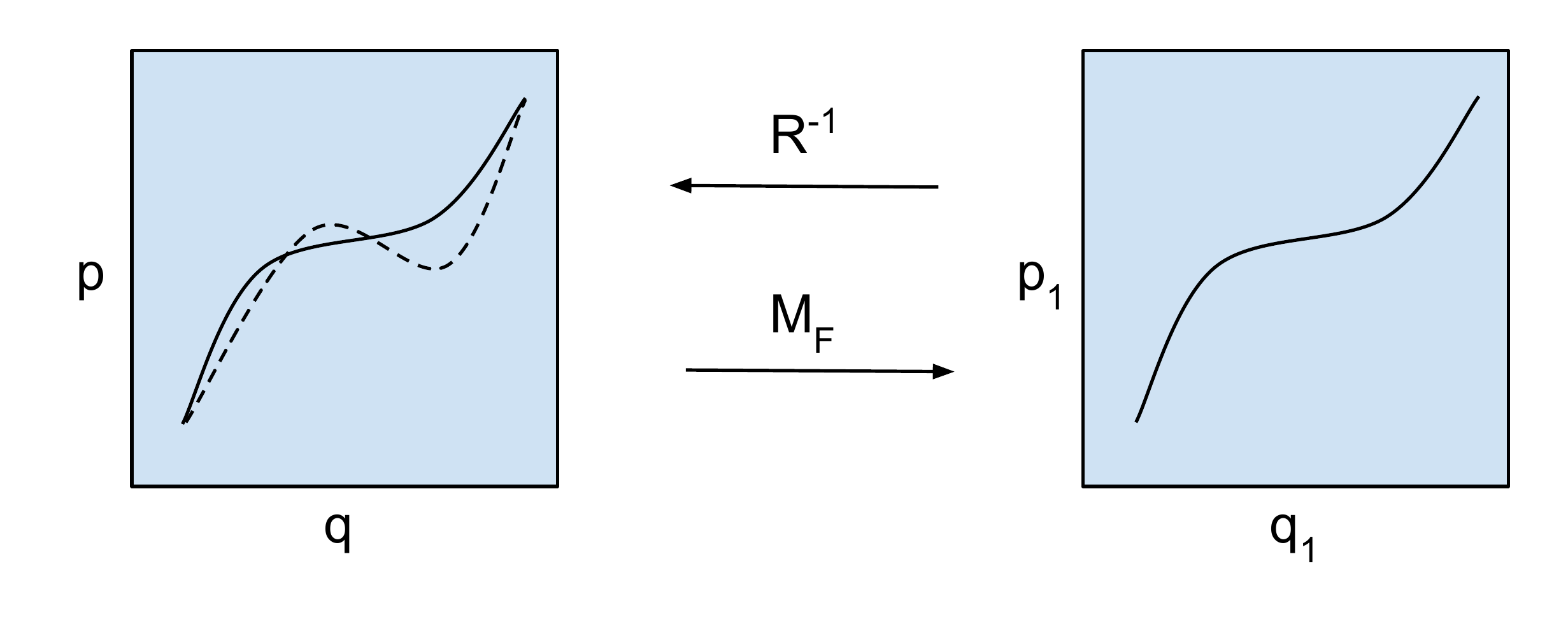}
\end{tabular}
\caption{\small  The discrepancy between formal canonical and coordinate map must lead to discrepancies between semiclassical descriptions. On the left we have an effective trajectory, which is mapped into a different time formulation on the right making use of the formal correspondence. Then one applies the inverse of coordinate map to compare the physical content of the semiclassical trajectory on the right with the original one on the left (for simplicity we suppress the time dimension). We assume that the map $\mathcal{M}_F$ is time-independent (which can always be achieved with appropriate choice of basic variables) so that the true hamiltonian transforms like observables in (\ref{trans}).}
\label{fig6}
\end{figure}

There are three immediate consequences:

\noindent {\bf I.} Suppose we set a semiclassical state as the initial condition for the unitary evolution starting in region I. Then the evolution of the state across region II will depend on the choice of time function there. One may determine semi-classical trajectories of the evolving cosmological system. These trajectories can be then used to determine the geometry of the bouncing universes. The bouncing scenarios must depend on the choice of time in a way explained in Fig \ref{fig6}.

\noindent {\bf II.} Since the clocks $X$ and $\tilde{X}$ are the same in region I and III, for the observer who registers only the initial (region I) and the final (region III) state of the universe and for whom the details of the intermediate evolution are unknown, this observer has no way to decide according to which clock the evolution has proceeded. In other words, the corresponding Schr\"odinger equations start out in the same form in region I, disagree in region II to become later one and the same equation again in region III. The same holds true for the quantum state evolved with those clocks from the same initial wavefunction. This suggests limits on our knowledge of the very early Universe.

\noindent {\bf III.} In cosmological applications, one is interested in incorporating the cosmological perturbations on the background geometry. In bouncing scenarios, one often seeks to challenge the inflationary paradigm by finding a way for just the right number of particles being produced due to the bounce. But, as noted before, the bounce is the intermediate phase, which is sensitive to the choice of time function. Therefore, at least within the usual approach based on effective semiclassical background metric, the power spectra of primordial perturbations will be sensitive to that choice.

\subsection{Example}\label{D}
Suppose that the time function is real, the space of Dirac observables is the plane and the constraint surface is parametrized as follows:
\begin{equation}
(q,p,X)\in \mathbb{R}^3,~\omega=\ud q\ud p ~~\Rightarrow~~\{\cdot,\cdot\}\big|_X=\frac{\partial\cdot}{\partial q}\frac{\partial\cdot}{\partial p}-\frac{\partial\cdot}{\partial p}\frac{\partial\cdot}{\partial q},
\end{equation}
and the true hamiltonian $h = 0$. We introduce a new time function (which is linearly related to the old one) and new basic variables via a coordinate transformation, $\mathcal{R}$:
\begin{equation}\label{pass}
\mathcal{R}:~(X,q,p)\mapsto (X_1\equiv X-q\in \mathbb{R},~~p_1\equiv p\in \mathbb{R},~~q_1\equiv X\in \mathbb{R}),\end{equation}
in which the reduced two-form reads:
\begin{equation}
\mathcal{R}^{-1*}\omega=\ud (X_1+q_1)\ud p_1=\ud q_1\ud p_1-\ud X_1\ud h_1
\end{equation}
where $h_1=p_1$. The above (passive) transformation $\mathcal{R}$ is of course not canonical as:
\begin{equation}
\{q_1,p_1\}\big|_X\equiv \{X,p\}\big|_X=0
\end{equation}
To have a canonical transformation we need to (actively) map constant-$X$ submanifolds into constant-$X_1$ submanifolds, i.e. map the Poisson brackets as well. Such a transformation reads:
\begin{equation}
\mathcal{M}_F: (X,q,p)\mapsto (X_1:=F(X),~q_1:=q+F(X),~p_1:=p)
\end{equation}
for some monotonic $F$ and it follows that
\begin{equation}
\mathcal{M}_F|_{X}: \{q,p\}_{X}\mapsto \{q_1-X_1,p_1\}_{X_1=F(X)}=1
\end{equation}
The transformation is active because in addition to the change of variables it maps physically inequivalent Poisson structures into each other. The transformation $\mathcal{M}_F$ is not unique in the sense that if combined with any active canonical transformation in the plane $(q,p)$, it gives another canonical transformation between these two systems, $\mathcal{M}_F'$. However, $\mathcal{M}_F$ is unique in the sense that it satisfies $\pi\circ \mathcal{M}_F=\pi$, i.e. the condition, which is necessary if we want to apply the map to the study of the multiple choice problem. From now on we assume $F(\cdot)=Id(\cdot)$ and drop $F$ from $\mathcal{M}_F$. When we quantize the two deparameterizations we assign quantum operators to the basic variables:
\begin{equation}\label{q0}
q\mapsto\hat{q}:=x,~~p\mapsto\hat{p}:=\frac{1}{i}\frac{d~}{d x},
\end{equation}
acting in $L^2(\mathbb{R},\ud x)$, and 
\begin{equation}\label{q1}
q_1\mapsto\hat{q}_1:=x_1,~~p_1\mapsto\hat{p}_1:=\frac{1}{i}\frac{d~}{d x_1},
\end{equation}
acting in $L^2(\mathbb{R},\ud x_1)$, and we quantize accordingly some relevant compound observables defined on the respective Hilbert spaces. Then, the existence of a unitary map $\mathbb{M}|_X$ corresponding to the active canonical map $\mathcal{M}|_{X}$ ensures that both quantizations are based on the same quantum representation of Dirac observables. The unitary transformation $\mathbb{M}|_X$ gives only formal correspondence between those two theories and one also needs the map (\ref{pass}) to make the {\it physical} connection between them. $\mathbb{M}|_X$ satisfies:
\begin{equation}
\mathbb{M}\hat{q}\mathbb{M}^{\dagger}=\hat{q}_1-X_1,~~\mathbb{M}\hat{p}\mathbb{M}^{\dagger}=\hat{p}_1
\end{equation}
from which it follows that (up to a phase factor)
\begin{equation}\label{u}
\mathbb{M}: L^2(\mathbb{R},\ud x)\ni \psi(x)\mapsto\psi (x_1-X_1)\in L^2(\mathbb{R},\ud x_1)
\end{equation}
Now, making use of (\ref{tru_ham}) one finds that the true hamiltonian associated with the other deparameterization reads:
\begin{equation}
h\mapsto\hat{h}=-i\mathbb{M}\frac{d\mathbb{M}^{\dagger}}{dX_1}=\frac{1}{i}\frac{d~}{dx_1}=\hat{p}_1
\end{equation}
We note that $\mathcal{M}$, and consequently $\mathbb{M}$, is very simple in this case. However, the theorems are general and thus applicable to non-trivial cases as well. The usefulness of $\mathbb{M}$ lies in the fact that it also relates compound observables, which often may have many self-adjoint realizations,
\begin{equation}
\mathbb{M}\hat{O}(X,q,p)\mathbb{M}^{\dagger}=\hat{O'}(X_1,q_1,p_1).
\end{equation}
However, the interpretation of  $O(X,q,p)$ and $O'(X_1,q_1,p_1)$ are not the same and to make a comparison one needs the {\it coordinate relation} given by formula (\ref{pass}). One finds that $O'=O'(q-X,X,p)$, which must be different than ${O}(X,q,p)$. In other words, if we use two different time frameworks, to the same classical observable there will correspond a different quantum operator. To see how different, let us consider an example. The observable
\begin{equation}
O=p^2+X^2
\end{equation}
when quantized according to (\ref{q0}) it becomes
\begin{equation}\label{O0}
\hat{O}=-\frac{d^2}{dx^2}+X^2
\end{equation}
In the other parametrization, the same operator will correspond to another observable, namely
\begin{equation}
\mathcal{R}^*\circ\mathcal{M}^{-1*} O = p^2 + (q-X)^2=O+{\it q(q-2X)}
\end{equation}
Thus, due to the extra term $q(q-2X)$, it is another physical quantity, which is associated with the same operator, when viewed with another time function. On the other hand, one finds that the same physical variable $O$ reads in the other deparameterization as follows:
\begin{equation}
\mathcal{R}^{-1*}{O}=p_1^2+q_1^2
\end{equation}
and when quantized according to (\ref{q1}) it becomes
\begin{equation}\label{O1}
\hat{O}=-\frac{d^2}{dx_1^2}+x_1^2
\end{equation}
Let us compare the spectral properties of (\ref{O0}) and (\ref{O1}): with respect to $X$, observable $O$ has the continuous spectrum:
\begin{equation}
sp(\hat{O})\bigg|_X=(X^2,\infty)
\end{equation}
whereas with respect to $X_1$ it has the discrete spectrum:
\begin{equation}
sp(\hat{O})\bigg|_{X_1}=2n+1, ~~n=0,1,\dots
\end{equation}
Moreover, the first spectrum is doubly-degenerate while the other one is non-degenerate. In the first case, the eigenvectors are not square-integrable while in the other one they are. This is a profound change in the character of operator based on completely different resolutions of identity \cite{QM}. What they share are the following classical-level features: positivity and unboundedness. Therefore, when one speaks of operators in quantum gravity one must always refer to a specific time function.

\section{Semiclassical description}
We have already proved that semiclassical description depends on the choice of time function but we have not yet determined the extent of the unavoidable discrepancies. This is the goal of the present section. We make use of the example of the previous section. Namely, we assume the Dirac space to be the plane with real basic variables $(q,p)$ and time function $X$ to be real. We generate a whole bunch of new deparameterizations by assuming new time functions $t=X+f(q)$, where $f$ is arbitrary. This is perhaps the simplest redefinition of time function one can come up with, which nevertheless is general enough for our purposes. Note that the pace of clock is conserved under such a redefinition and so is the lapse function, $N_t=N_X\frac{dX}{dt}=N_X$. 

We use the Schr\"odinger representation given in (\ref{q0}) and we pick the gaussian state $\psi(x)=\pi^{-\frac{1}{4}}e^{-\frac{x^2}{2}}$. In what follows we study the semiclassical dynamics of observables of the form $\mathcal{O}=\mathcal{O}(q,X)$, i.e. for simplicity we suppress the dependence on $p$. Such observables can be approximated by polynomials:
\begin{equation}
\mathcal{O}=\sum_{n,m} O_{nm}q^nX^m
\end{equation}
where $O_{nm}$ are constants. We focus on a single term of the above sum, and find an expectation value for it, with respect to $X$ and to $t$
\begin{equation}\label{semi0}
\langle q^nX^m\rangle = \begin{cases}\int_{-\infty}^{\infty} \pi^{-\frac{1}{2}}e^{-x^2}x^nX^m dx &\mbox{wrt X}\\ \int_{-\infty}^{\infty} \pi^{-\frac{1}{2}}e^{-x^2}x^n(t-f(x))^m dx &\mbox{wrt t}\end{cases}
\end{equation}
where we employed the coordinate map to determine the corresponding operator in time `$t$'. Now, let us set $f(x)=x^{k},~k\in \mathbb{Z}$. We find
\begin{equation}\label{semi}
\int_{-\infty}^{\infty} \pi^{-\frac{1}{2}}e^{-x^2}x^n(t-x^k)^m dx=\sum_{l=0}^{m} (-1)^l t^{m-l} \frac{m!(n+lk)!}{l!(m-l)!2^{\frac{n+lk}{2}}(\frac{n+lk}{2})!}~\sigma_{n+kl}
\end{equation}
where $\sigma_s=1$ if $s$ is even and vanishes otherwise. We may safely fix $n=0$ since $n$ regulates the dependence on Dirac's observable only.

Let us study the linear case $m=1$. We have $\langle X\rangle(X)=X$ and
\begin{equation}
\langle X\rangle(t) = \begin{cases} t-\frac{k!}{2^{\frac{k}{2}}(\frac{k}{2})!} &\mbox{for k even} \\ 
t & \mbox{for k odd}. \end{cases} 
\end{equation}
from which we see that choosing another time function may only shift the original time dependence by a constant.

Let us study the quadratic case $m=2$. We have $\langle X^2\rangle(X)=X^2$ and
\begin{equation}
\langle X^2\rangle(t) = \begin{cases} t^2-\frac{2k!}{2^{\frac{k}{2}}(\frac{k}{2})!} t+\frac{(2k)!}{2^kk!} &\mbox{for k even} \\ 
t^2+\frac{(2k)!}{2^kk!} & \mbox{for k odd}. \end{cases} 
\end{equation}
from which we see that choosing another time function may change the original time dependence in such a way that the minimum of the dynamical observable is lifted. We find that
\begin{eqnarray}
\min \langle X^2 \rangle(X)&=&0,\\
\min \langle X^2 \rangle(t)&=&\frac{(2k)!}{2^kk!}-\left(\frac{k!}{2^{\frac{k}{2}}(\frac{k}{2})!}\right)^2~~\textrm{$k$ is even},\\
\min \langle X^2 \rangle(t)&=&\frac{(2k)!}{2^kk!}~~\textrm{$k$ is odd}.
\end{eqnarray}
We notice that by changing time function wrt which measurements are made, one may lift the minimum as much as one wishes - both for $k$ odd and $k$ even, by increasing $k$. The quadratic case is particularly interesting because it may model for instance the semiclassical behavior of volume or the inverse of energy density in quantum cosmological scenarios at the big bounce. We have found that the minimal volume of the universe or the maximal value of energy density in such scenarios will depend on the chosen time. This example shows that the relevance of Planck scale is a subtle issue in quantum gravity, when one considers dynamical observables. Actually, we have proved that there is no fundamental scale invariably associated with a dynamical observable.

Our method does not involve true hamiltonians. For the deparameterizations that we consider true hamiltonians vanish and the dynamics is completely contained in the time parameters $X$ and $t$. This, however, is due to the specific choice of basic variables. A parallel analysis could be made with another choice of basic variables and non-vanishing true hamiltonians\footnote{As far as there exists a unitary transformation corresponding to the employed canonical one.}. In other words, the choice of time function and the form of true hamiltonian are unrelated and we used this fact to make our analysis simple and transparent. We wish to stress that the space of observables as well as the transformations of time functions which we considered here are very limited and more general considerations may only increase the extent of the multiple choice problem.

\begin{figure}[t]
\begin{tabular}{cc}
\includegraphics[width=0.45\textwidth]{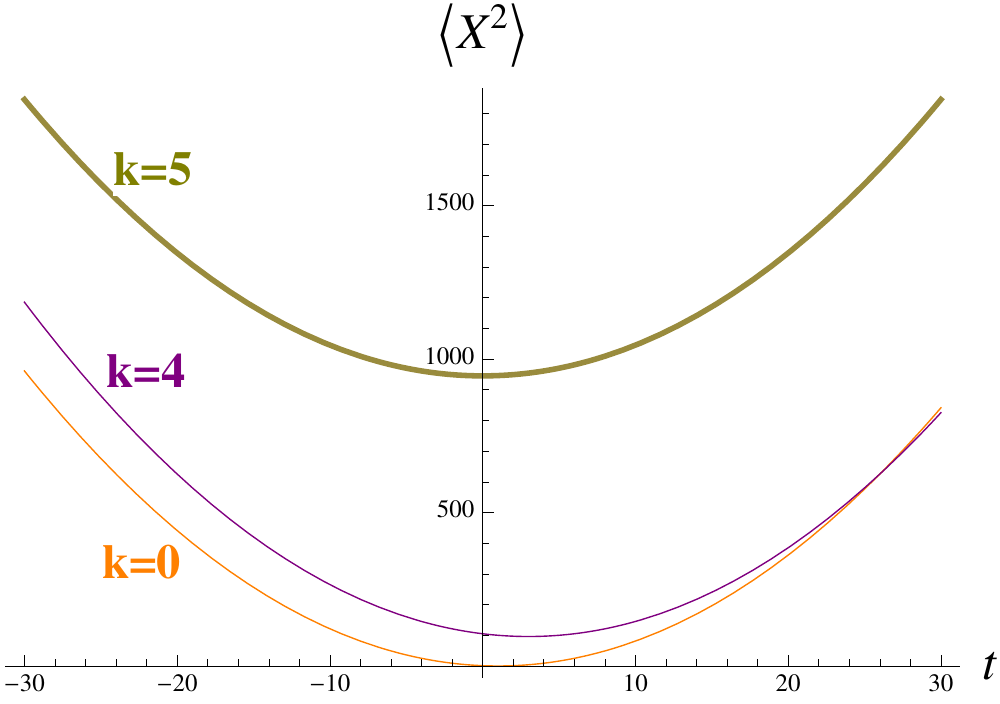}
\end{tabular}
\caption{\small  The behavior of $\langle X^2\rangle$ with respect to different time functions. We fix an observable by setting $n=0,~m=1$ (see formula (\ref{semi0})), while time functions are picked by setting $k=0,4,5$ (see formula (\ref{semi})). The minimal value and the moment of bounce are shifted. All these effective trajectories are obtained from the same semiclassical state in the carrier space of the representation of Dirac observables.}
\label{fig7}
\end{figure}   

\section{Discussion}
To obtain a quantum description of a gravitational system, a choice of time function must be made. The understanding of the relation between quantum descriptions with different time functions is necessary if we are to have a reliable quantization procedure. In this article we made a first step toward this goal by proving a formal, as opposed to physical, canonical relation between different time-based canonical descriptions. We managed to separate results of the different choice of time function from results of inequivalent quantizations. This circumstance let us prove that spectra of quantized dynamical observables as well as semiclassical trajectories will depend on employed clocks. Both facts were illustrated with clear examples. From these facts a few important, though rarely acknowledged, consequences follow. First of all, it is meaningless to talk about spectrum of any operator in quantum gravity unless one refers to a specific choice of time function. In this light, spectra of the so called kinematical operators (before the imposition of constraints) are weak prognostics for the spectra of corresponding physical operators (after the imposition of constraints). Analogously, a semiclassical description must not be trusted as it is only one realization of multiple scenarios which come with {\it any fixed} quantization of Dirac observables. In particular, statements regarding maximal energy density or quantities of other physical dimensions at the moment of big bounce cannot be meaningful without a reference to a specific time function. In the forthcoming article, we should determine the implication of the choice of time for specific quantum cosmological models.

It seems that the theory of cosmological perturbations may provide an interesting setting for the multiple choice problem investigations. The particle production in the inflationary phase is successfully used to set the initial conditions for matter distribution in the early Universe. However, it might also be possible to model the creation of primordial perturbations in a similar manner during the big bounce phase \cite{peter}. For that purpose, if the bounce is due to quantum gravity effects, one employs a framework, in which both the background metric and the cosmological perturbations are quantized. Then the usual quantum theory on curved spacetimes is attained after a kind of semiclassical background geometry is assumed (see e.g. \cite{lewand}). Since the outcome of the whole process is induced by semiclassical background geometry, it must be choice-of-time-dependent. In effect, the CMB power spectra may provide some hints into the time problem in the early universe. 

Let us comment on some related current works on the multiple choice problem. In \cite{Dittrich:2007th} authors develop an argument against the existence of a relation between the spectral properties of kinematical and some of the ``corresponding" physical observables. They emphasize the reduction of Hilbert space through a constraint operator equation, which spoils the alleged relation. This interesting work thus provides a complementary viewpoint on the problem. In our setup we have no information on the initial kinematical phase space and the spectral properties of physical operators are shown to vary due to the lack of a God-given clock. We emphasize the lack of predetermined Poisson structure in the constraint surface.

The recent developments by M. Bojowald and his collaborators \cite{Boj} are particularly interesting in the context of our results. Their goal is a semiclassical description of totally constrained systems, which do not necessarily admit a global clock. Their method consists of the Dirac quantization followed by solving the semi-classical constraint with a local choice of clock. They find that the semiclassical evolution in different local clocks can be matched for all observables up to second order in basic variables (higher orders are not investigated). This seemingly contradictory result may in fact be providing a complementary picture of what is going on at the level of semiclassical Hamiltonian constraint and it should be very interesting to have a clear comparison of these two approaches.

\acknowledgments This work was supported by Narodowe Centrum Nauki with decision No. DEC-2013/09/D/ST2/03714. I thank Edward Anderson, Herv\'e Bergeron and Martin Bojowald for their comments.

\end{document}